\documentclass[acmsmall]{acmart}
\usepackage{xspace}
\usepackage{caption}
\usepackage{xcolor}  
\usepackage{subcaption}
\AtBeginDocument{%
  }

\setcopyright{acmlicensed}
\copyrightyear{2018}
\acmYear{2018}
\acmDOI{XXXXXXX.XXXXXXX}

\acmJournal{JACM}
\acmVolume{37}
\acmNumber{4}
\acmArticle{111}
\acmMonth{8}

\newcommand{\yzz}[1]{\textcolor{magenta}{#1}}

\usepackage{xcolor}
\usepackage{enumitem}

\usepackage{booktabs}
\usepackage{pifont}
\usepackage{graphicx}
\usepackage[skins]{tcolorbox}
\usepackage{array}
\usepackage{tabularx}
\usepackage{booktabs}
\usepackage{siunitx}
\usepackage{rotating}
\usepackage{soul}

\newcommand{\finding}[2]{%
  \begin{tcolorbox}[
    enhanced,
    colback=gray!10,
    colframe=black,
    boxrule=0.8pt,
    arc=1pt,
    left=6pt, right=6pt, top=6pt, bottom=6pt
  ]
    \textbf{Finding #1.} \small #2
  \end{tcolorbox}%
}

\usepackage{colortbl}
\usepackage{xcolor}
\usepackage{booktabs}

\definecolor{heat0}{gray}{1.00}    
\definecolor{heat1}{gray}{0.95}    
\definecolor{heat2}{gray}{0.88}    
\definecolor{heat3}{gray}{0.80}    
\definecolor{heat4}{gray}{0.70}    
\definecolor{heat5}{gray}{0.60}    
\definecolor{heat6}{gray}{0.50}    
\definecolor{heat7}{gray}{0.40}    
\definecolor{heat8}{gray}{0.30}    

\newcommand{\heatcell}[1]{%
  \ifnum#1=0 \cellcolor{heat0}#1%
  \else\ifnum#1<6 \cellcolor{heat1}#1%
  \else\ifnum#1<16 \cellcolor{heat2}#1%
  \else\ifnum#1<51 \cellcolor{heat3}#1%
  \else\ifnum#1<151 \cellcolor{heat4}#1%
  \else\ifnum#1<501 \cellcolor{heat5}#1%
  \else\ifnum#1<1001 \cellcolor{heat6}#1%
  \else\ifnum#1<2001 \cellcolor{heat7}\textcolor{white}{#1}%
  \else \cellcolor{heat8}\textcolor{white}{#1}%
  \fi\fi\fi\fi\fi\fi\fi\fi%
}

\newcommand{\heatcellL}[2]{
  \ifnum#1=0 \cellcolor{heat0}#2%
  \else\ifnum#1<6 \cellcolor{heat1}#2%
  \else\ifnum#1<16 \cellcolor{heat2}#2%
  \else\ifnum#1<51 \cellcolor{heat3}#2%
  \else\ifnum#1<151 \cellcolor{heat4}#2%
  \else\ifnum#1<501 \cellcolor{heat5}#2%
  \else\ifnum#1<1001 \cellcolor{heat6}#2%
  \else\ifnum#1<2001 \cellcolor{heat7}\textcolor{white}{#2}%
  \else \cellcolor{heat8}\textcolor{white}{#2}%
  \fi\fi\fi\fi\fi\fi\fi\fi%
}

\begin{document}

\title{MOA: A Profiling-Guided LLM Framework for Memory-Optimization Automation at Codebase Scale}

\author{Jiaxi Liang}
\authornote{Both authors contributed equally to this research.}
\email{jiaxi.liang@connect.hku.hk}
\affiliation{%
  \institution{The University of Hong Kong}
  \city{Hong Kong}
  \country{China}
}

\author{Yuanxiang Shi}
\email{shiyuanx@hku.hk}
\authornotemark[1]
\affiliation{%
  \institution{The University of Hong Kong}
  \city{Hong Kong}
  \country{China}
}

\author{Zezhou Yang}
\email{zezhouyang@connect.hku.hk}
\affiliation{%
  \institution{The University of Hong Kong}
  \city{Hong Kong}
  \country{China}
}

\author{Chenxiong Qian}
\email{cqian@cs.hku.hk}
\authornote{Corresponding Author.}
\affiliation{%
  \institution{The University of Hong Kong}
  \city{Hong Kong}
  \country{China}
}





\renewcommand{\shortauthors}{Trovato et al.}

\begin{abstract}

Modern large-scale software systems often suffer from pervasive memory inefficiencies (e.g., bloat, churn), leading to excessive resource costs and performance degradation.
Existing optimization workflows lack end-to-end automation, forcing developers to manually synthesize complex tool outputs into actionable and semantics-preserving fixes, precluding scalability in large codebases.
To address this, this paper presents \sys, an LLM-driven framework that automatically detects and repairs recurring memory inefficiencies across production-scale codebases. 
Specifically, \sys operates through three agents: an Analyzer that mines anti-patterns from profiling data, a Checker Generator that synthesizes static analyzers through template-guided refinement, and a Patcher that generates optimization patches via state-machine-driven workflows. 
Our evaluation on OpenHarmony, an open-source operating system with over 100 million lines of C/C++ code, shows that \sys identifies 13 anti-patterns~(9 previously unknown) from 3 profiled services, detects over 10,000 inefficiencies across a broader set of 7 services, and generates 769 patches with 92.5\% expert acceptance rate, achieving 42.2\% heap reduction and 10.6\% binary size reduction on average. 
We envision \sys as a valuable tool for performance engineering at production scale. 
\end{abstract}



\newcommand{\sys}{\textsc{MOA}\xspace}
\newcommand{\cmmnt}[1]{}

\maketitle

\section{Introduction}

As production software evolves, memory inefficiencies often accumulate as \textit{silent technical debt}, manifesting in memory bloat and degraded responsiveness that are difficult to diagnose and rectify automatically~\cite{jin2012understanding, 10.5555/2664446.2664477, 10.1145/2714064.2660234, 10.1145/2259016.2259033}. 
Unlike functional bugs that cause crashes or incorrect outputs, memory inefficiencies usually do not break program execution and therefore often remain unnoticed until deployment~\cite{10.5555/2664446.2664477, 10.1145/2714064.2660234, dai2017hytrace}. 
Consequently, identifying and rectifying these inefficiencies remains a daunting manual task, requiring developers to possess deep system-level expertise to bridge the gap between low-level symptoms and source-code root causes~\cite{jin2012understanding, 10.5555/2664446.2664477}.
This process is not only labor-intensive but also difficult to scale across production-grade codebases, emphasizing the urgent need for automated and systematic solutions.

Large Language Models (LLMs) have recently emerged as a promising avenue for such automation, having significantly expanded the scope of automated software engineering~\cite{10.1145/3747588, Jimenez2023-ve, Zhang2023-em, DBLP:journals/tosem/YangCGLHLX25}. However, applying LLMs to memory optimization is far from straightforward and faces several fundamental challenges:

\textit{C1: Large-scale software structures hinder the precise localization of memory inefficiencies.}
Memory inefficiencies are often workload-dependent and intertwined with complex control flow and data lifetimes, thus locating their root causes without runtime evidence is difficult. 
While profiling can reveal concrete symptoms, manually bridging profiling data to actionable insights remains slow and error-prone at scale~\cite{jin2012understanding, olivo2015static}, and raw profiling data cannot be directly leveraged by LLMs.

\textit{C2: Localized profiling symptoms do not directly transfer to codebase-wide detection.} 
Profiling captures specific instances under specific executions, but scalable detection requires recognizing the underlying recurring patterns and finding their occurrences across the codebase. 
Bridging this gap demands lifting symptoms into generalizable patterns and encoding them as static detection rules.

\textit{C3: Complex code dependencies make memory optimization difficult to carry out stably and consistently.}
Memory optimizations in codebases rarely reduce to isolated edits at a single reported location. 
Instead, they often require understanding surrounding code, tracking related symbol usages, and coordinating multiple changes across files or functions. 
As a result, even after an inefficiency has been identified, turning it into a stable and reliable code change remains challenging~\cite{nistor2015caramel, selakovic2015automatically}.

\textit{C4: Ensuring reliable and consistent memory optimization remains challenging due to the lack of direct validation oracles.}
Unlike tasks with clear pass-or-fail outcomes, memory optimization offers no direct oracle for judging whether detections or generated changes are reliable enough to trust~\cite{nistor2013toddler, pradel2014performance}.
This makes automated workflows dependent on additional mechanisms for assessing the quality and consistency of intermediate results.

\textbf{Approach.} 
To address these challenges, we propose \textbf{\sys}, a LLM-driven framework for \textbf{M}emory-\textbf{O}ptimization \textbf{A}utomation at codebase scale with three stages.

\begin{figure*}[t]
  \centering
  \includegraphics[width=\textwidth]{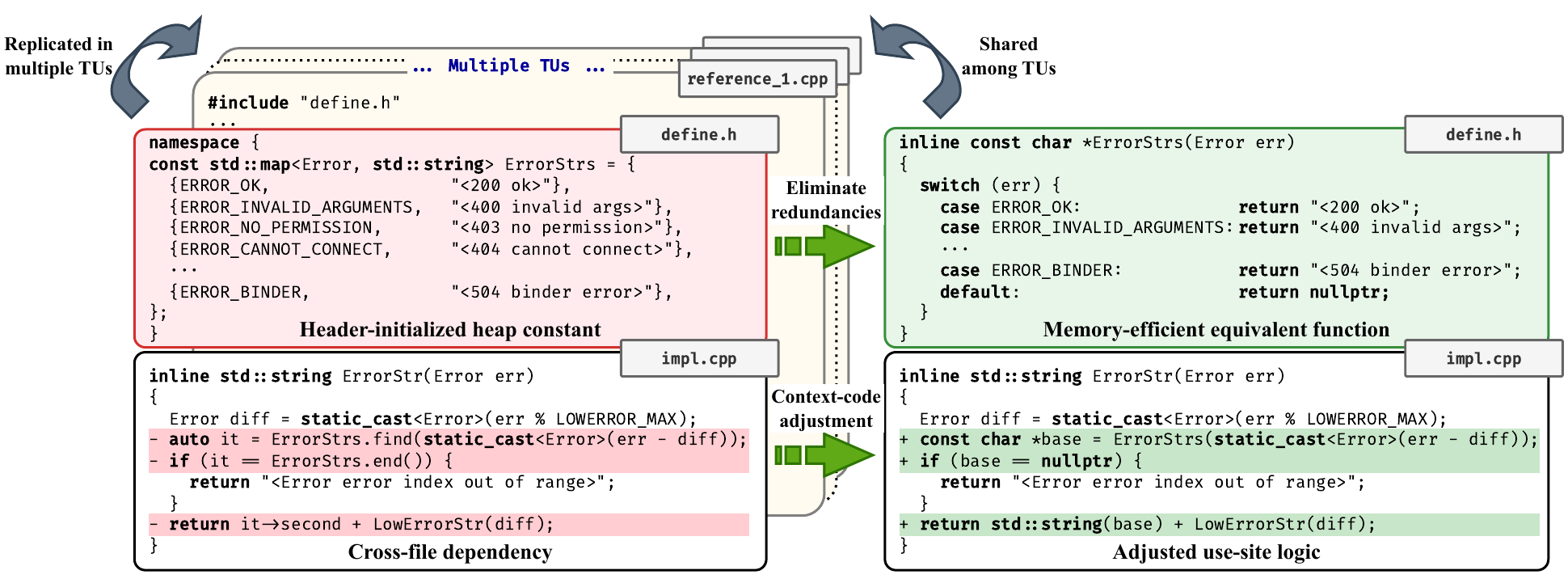}
  \caption{Motivating example of eliminating header-initialized constant map. \cmmnt{\yzz{Two observations should be presented in the figure.}}}
  \label{fig:motivating-example}
\end{figure*}

\textit{Stage 1: Pattern Mining.}  
In the first stage, reusable anti-patterns are extracted from specific instances of inefficiency identified from profiling data~(C1).
By importing execution traces into a database, an \textit{Analyzer} agent infers anti-patterns through profiling data analysis and source code exploration, generating candidate pattern reports~(C2).
An automated validation step then checks each candidate pattern against predefined criteria and provides refinement feedback, ensuring only verified anti-pattern reports proceed to checker synthesis~(C4).

\textit{Stage 2: Checker Synthesis.} 
Then, we design a \textit{Checker Generator} to synthesize executable static analyzers from validated anti-patterns, automating the complex pattern rule formalization process through iterative generation and refinement: prototype synthesis creates a compilable checker, then refinement progressively improves detection logic under self-validation~(C2).
This loop continues until the checker correctly identifies all test cases while eliminating false positives and negatives~(C4).

\textit{Stage 3: Patch Generation.} 
Finally, our approach automatically generates optimization patches to resolve the detected inefficiencies.
To handle large volumes of detected targets, a \textit{Patcher} agent first groups related targets into independent chunks, then employs a three-stage workflow that drives the agent through context gathering, edit generation, and validation, automating the complex fixing process~(C3).
Within the workflow, syntax checking and automatic state transitions enable iterative refinement until all inefficiencies are correctly patched~(C4).

\textbf{Evaluation.} 
We evaluate \sys on OpenHarmony ~\cite{10.1145/3720538}, an open-source operating system. 
Leveraging profiling data collected on 3 system services, \sys identifies 9 previously unknown anti-patterns, synthesizes 13 validated static checkers, detects 10,067 inefficiency instances across the codebase, and generates 769 optimization patches within 7 selected services. 
After being reviewed by human maintainers, 92.5\% of these patches are accepted as valid patches, achieving an average of 42.2\% heap size reduction and 10.6\% binary size reduction.

\textbf{Contributions.} 
In summary, the three main contributions of this paper are as follows:
\begin{itemize}[label=\large$\bullet$, topsep=1pt, leftmargin=2.5em]
    \item We propose \textbf{\sys}, a fully automated LLM-based framework for codebase-scale memory optimization. 
    By combining runtime profiling with LLM-driven analysis and transformation, \sys bridges the gap from localized inefficiency symptoms to codebase-wide detection and fixing.
    \item We conduct a \textbf{comprehensive evaluation} on OpenHarmony, a production-scale operating system. 
    \sys successfully validates the complete pipeline from pattern mining through checker synthesis to automated repair, demonstrating substantial memory and binary size reductions with high patch acceptance rates across hundreds of detected instances.
    \item To our knowledge, \sys is the \textbf{first autonomous framework} capable of detecting and repairing recurring memory inefficiencies across a production-scale codebase. 
    We will release our implementation to facilitate future research.
\end{itemize}
\section{Motivation}

\paragraph{A Real-World Case}
A pervasive memory anti-pattern in OpenHarmony arises from defining and initializing non-trivial static objects in header files.
Due to C++ internal linkage rules, such objects are instantiated independently in every translation unit (TU) that includes the header.
Figure~\ref{fig:motivating-example} illustrates a representative case: a static constant mapping table that is duplicated over 100 times, as revealed by our profiling. 
This redundancy triggers significant binary-size bloat and redundant heap allocations during program initialization. 
While replacing the \emph{std::map} table with an inline lookup function and substituting \emph{std::string} with character pointers is a conceptually straightforward optimization, carrying out such refactoring safely and in a semantics-preserving manner across a large codebase remains a daunting manual task.

\begin{figure*}[t]
  \centering
  \includegraphics[width=\textwidth]{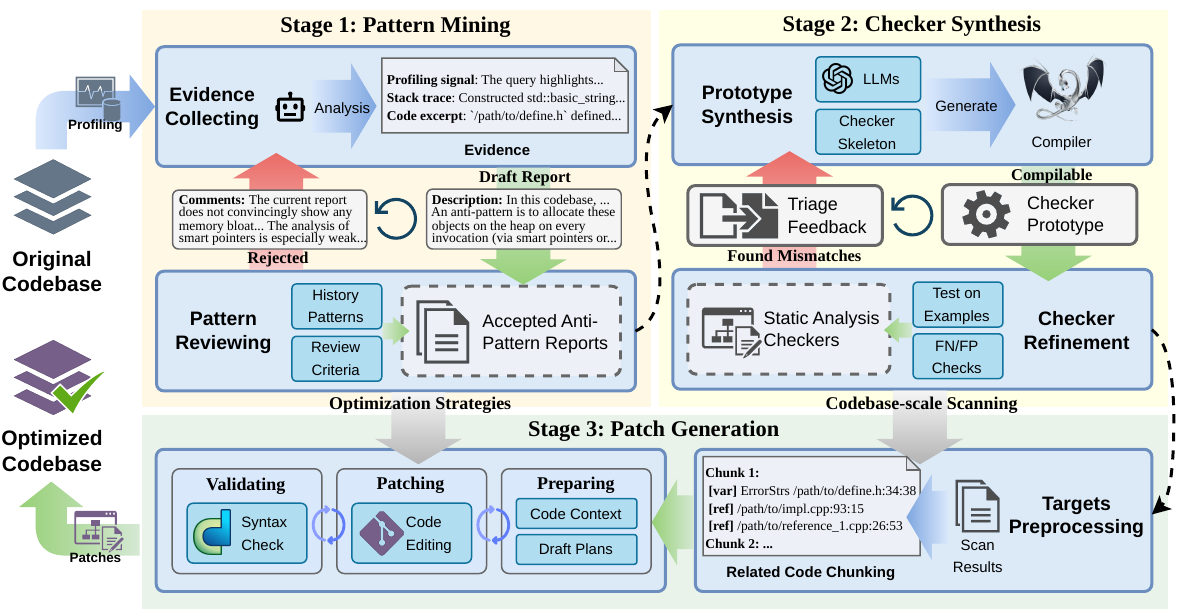}
  \caption{Overview of \sys. {\mdseries The process begins with the original codebase and its profiling data. (1)~In \textit{Pattern Mining}, \sys mines recurring anti-patterns from profiling evidence and code context. (2)~In \textit{Checker Synthesis}, it converts the mined reports into static checkers for codebase-wide detection. (3)~In \textit{Patch Generation}, the \textit{Patcher} groups detected targets into chunks and iteratively edits and validates patches. The validated patches are then applied to the codebase, producing an optimized codebase.}}
  \label{fig:overview}
\end{figure*}

\textit{Observation 1: Recurrence and Semantic Abstraction.} 
As shown in Figure~\ref{fig:motivating-example}, profiling may identify the mapping table \emph{ErrorStrs} as a memory hotspot.
However, the real issue is not this specific table itself, but the underlying anti-pattern: initializing non-trivial static objects in header files.
While profiling reveals the symptom of being replicated across multiple TUs, recognizing it as a recurring inefficiency requires lifting this concrete instance into a semantic rule.
An automated approach must infer that any object exhibiting this anti-pattern can introduce redundant heap allocations and binary duplication across the codebase, thereby moving from instance-level symptoms to pattern-level abstraction.

\textit{Observation 2: Coordinated and Context-dependent Repair.}
The transition from the inefficient version to the optimized version in Figure~\ref{fig:motivating-example} shows that memory optimization is rarely a purely local fix.
The required context-code adjustment involves coordinated modifications: in this example, the definition in \emph{define.h} is refactored into a lookup function, while call sites in files such as \emph{impl.cpp} are updated accordingly to handle the changed interface.
Such non-atomic changes require a broader understanding of program logic across files to ensure that the refactoring remains semantics-preserving and does not introduce regressions in dependent code.

These observations suggest that effective automation should focus on higher-level anti-pattern abstraction rather than directly addressing isolated profiling instances, and should support coordinated, context-aware code changes rather than local edits alone.
To operationalize these insights, a framework must possess three core capabilities: (1) extracting generalizable anti-patterns from dynamic runtime symptoms, (2) scaling detection across the codebase, and (3) synthesizing context-sensitive patches. 
In the following section, we detail how \sys fulfills these requirements through a structured LLM-driven pipeline.

\section{The \sys Design}

\subsection{Framework Overview}

Figure~\ref{fig:overview} illustrates the overall workflow of \sys. 
Given a dynamic profiling report and the original codebase, \sys uses three stages to automate memory optimization. 
First, the \emph{Pattern Mining} stage correlates profiling evidence with program semantics to extract recurring memory anti-patterns and summarize them as structured reports.
Subsequently, the \emph{Checker Synthesis} stage converts these reports into static analysis checkers, enabling scalable, codebase-wide detection of similar inefficiencies.
Finally, the \emph{Patch Generation} stage takes the checker-detected targets together with the optimization strategies provided by anti-pattern reports, and produces semantics-preserving optimization patches.
By systematically integrating these stages, \sys transforms transient runtime symptoms into persistent, actionable optimizations, effectively bridging the gap between localized profiling and codebase-wide optimizations.

\subsection{Pattern Mining}

The first stage analyzes profiling results to identify recurring memory anti-patterns.
We build the \textit{Analyzer}, an LLM-based agent that flexibly explores profiling results to mine anti-patterns.
Specifically, \sys examines fine-grained runtime events and execution traces, together with source-level code information, to extract evidence from concrete inefficiency symptoms and infer structured anti-pattern reports.

\paragraph{Evidence Collecting}
To support flexible evidence exploration over large profiling traces, we import processed profiling data into a relational database and expose it to the \textit{Analyzer} through a structured query interface.
The database provides a bounded yet flexible environment for evidence exploration: instead of loading full profiling reports into context, the agent retrieves only relevant slices of data and incrementally analyzes them under different views and granularities. 
This design keeps large traces manageable while making the exploration process reproducible, auditable, and constrained to well-defined database operations.

Profiling data alone does not fully explain the source-level semantics behind runtime symptoms.
We therefore allow the \textit{Analyzer} to inspect the target codebase during analysis.
Starting from profiling evidence, the \textit{Analyzer} first infers candidate anti-patterns, then examines relevant source code to confirm their manifestations and understand their semantic causes, and finally summarizes its findings into structured anti-pattern reports. 
By grounding pattern inference in both runtime observations and source-level analysis, this process links profiling symptoms to program semantics and improves the reliability of the resulting reports.

\begin{figure}[t]
  \centering
  \includegraphics[width=0.6\columnwidth]{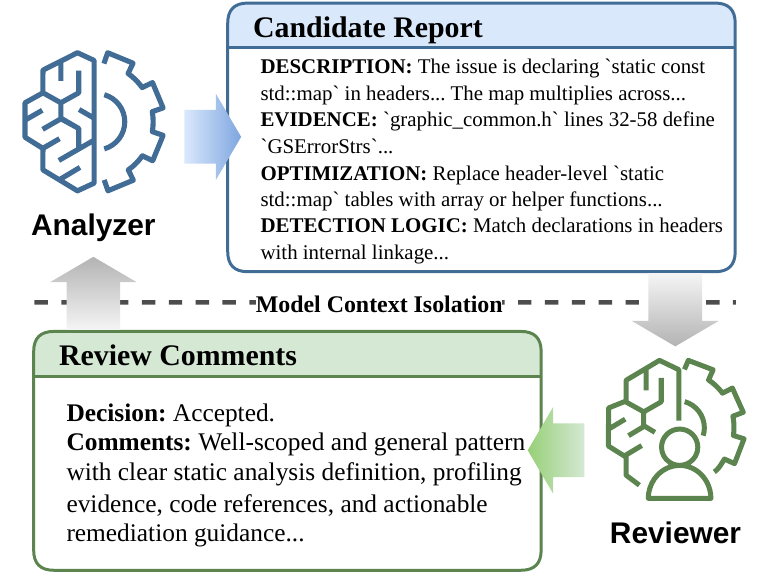}
  \caption{Example of report drafting \& reviewing iteration.}
  \label{fig:analyzer-overview}
\end{figure}

\paragraph{Report Drafting}
Following the evidence exploration, the \textit{Analyzer} formulates a structured anti-pattern report, serving as the pivotal intermediary that directs both the static checker synthesis and the patch generation workflows.
To ensure its quality and consistency, we define a formatted anti-pattern report template with four essential components: 
(1) a pattern description with evidence linking profiling symptoms to code examples; 
(2) an explanation of why the pattern causes inefficiency; 
(3) detection logic that can be translated into static analysis rules; and 
(4) actionable optimization strategies.
This structured template enforces consistency across components: the detection logic is grounded in empirical evidence to reduce false positives, while the optimization strategies are aligned with the diagnosed causes of inefficiency.
By imposing these structural constraints, \sys ensures that mined anti-pattern reports are internally consistent and usable by downstream stages.

\paragraph{Pattern Reviewing}

Structured templates alone are insufficient to guarantee that mined anti-pattern reports are suitable for downstream checker synthesis and patch generation. 
In practice, failures arise either from weak report formulation or from flaws in the proposed anti-pattern itself, indicating that quality control must go beyond template compliance.
We attribute part of this issue to a broader limitation of long-context LLM reasoning: once the \textit{Analyzer} has committed to a candidate explanation, it may struggle to critically reassess that result within the same context. 
Prior work reports similar self-reinforcing behavior in iterative LLM refinement~\cite{LLM-self-bias}. This motivates an external review step. 

To address this, \sys introduces an independent LLM-based Reviewer as a final quality gate for pattern mining. 
Given a candidate report and a review criteria prompt, the Reviewer independently examines whether the proposed anti-pattern is well supported, whether it overlaps with previously accepted patterns, and whether the report includes enough concrete evidence and code examples.
Based on this assessment, it returns a decision with comments indicating whether the report should be accepted or rejected.
This process ensures that validated anti-pattern reports provide a unique anti-pattern, representative examples, and actionable optimization guidance.
Through repeated analysis and review, the \textit{Analyzer} distills profiling evidence into reliable and actionable anti-pattern reports. 
These reports serve as the foundation of \sys, providing the pattern abstractions, examples, and optimization guidance on which subsequent detection and patch generation depend.

\subsection{Checker Synthesis}

Building upon the validated anti-pattern reports, the framework transitions from localized insights to codebase-scale detection by identifying latent instances that mirror the defined inefficiencies.
Static analyzers provide a practical substrate for this purpose, as they can encode recurring anti-patterns as reusable detection logic and apply that logic at scale across the codebase. 
This transition from abstract description to scalable execution is facilitated by a mechanism capable of mapping the reports' structured guidance, detection logics, and code exemplars into functional checker implementations. 
For this purpose, we design the \textit{Checker Generator}, which synthesizes customized static analysis checkers from anti-pattern reports through two phases: prototype synthesis and checker refinement.

\paragraph{Prototype Synthesis.}
The first phase focuses on the systematic transformation of an anti-pattern report into an  initial compilable checker prototype.
In our implementation, we instantiate this design on the Clang Static Analyzer (CSA), which provides a practical substrate for path-sensitive analysis and codebase-wide scanning in C/C++ projects.
LLMs can interpret the structured guidance in anti-pattern reports and map it to checker code, but this process is not always reliable: the generated code may violate framework-specific APIs, checker structure, or compilation requirements.
To make synthesis more stable, we constrain generation to pattern-specific logic within a predefined checker skeleton.
The generated checker is then compiled automatically, and compiler feedback is used to iteratively repair synthesis errors until a compilable prototype is obtained. 
In this way, prototype synthesis turns high-level anti-pattern reports into basic executable checker implementations.

\paragraph{Checker Refinement.}
A merely compilable checker is not necessarily a correct one. 
The successful synthesis of a compilable prototype does not inherently guarantee functional accuracy or semantic alignment with the target anti-pattern. 
Such checkers may still exhibit discrepancies in the form of false negatives or false positives, which necessitates a dedicated validation stage to ensure diagnostic reliability.

To achieve this, the \textit{Checker Generator} uses code examples extracted from the anti-pattern report as feedback cases. 
These examples provide concrete anchors for the intended detection behavior and help expose mismatches between the report’s abstract description and the checker’s actual implementation, such as false negatives on true pattern instances or false positives on superficially similar but irrelevant code.
The checker is executed on these examples, and the resulting outputs are compared against the expected behavior implied by the report. 
Such mismatches reveal underlying limitations in the checker’s completeness or precision, prompting the agent to revise the detection logic and re-initiate the synthesis process within an iterative refinement loop.

The culmination of this iterative cycle is the production of validated checkers that are precisely aligned with the mined anti-pattern specifications. 
By leveraging these specialized implementations, \sys extends its diagnostic reach beyond the constraints of the initial profiling data to identify recurring inefficiencies at codebase scale.

\subsection{Patch Generation}

The systematic identification of inefficiencies across the codebase via synthesized checkers establishes the necessary foundation for the subsequent application of optimization strategies to the detected targets.
Given that individual detection sites typically serve as entry points to broader architectural contexts rather than isolated fix locations, we introduce the \textit{Patcher}, an LLM-based agent that transforms diagnostic findings into stable, semantics-preserving optimization patches through the comprehensive analysis of surrounding dependencies and the coordination of distributed modifications.

\paragraph{Target Preprocessing.}

Codebase-wide scan results often require preprocessing before they can serve as effective patching units, as they are typically too numerous and redundant to be used directly as patching tasks.
Moreover, a reported detection location does not always provide enough context for deciding how the inefficiency should be addressed. 
Many memory optimizations involve related code sites, symbol usages, or surrounding implementation details that must be considered together.

\begin{figure}[htbp]
  \centering
  \includegraphics[width=0.63\columnwidth]{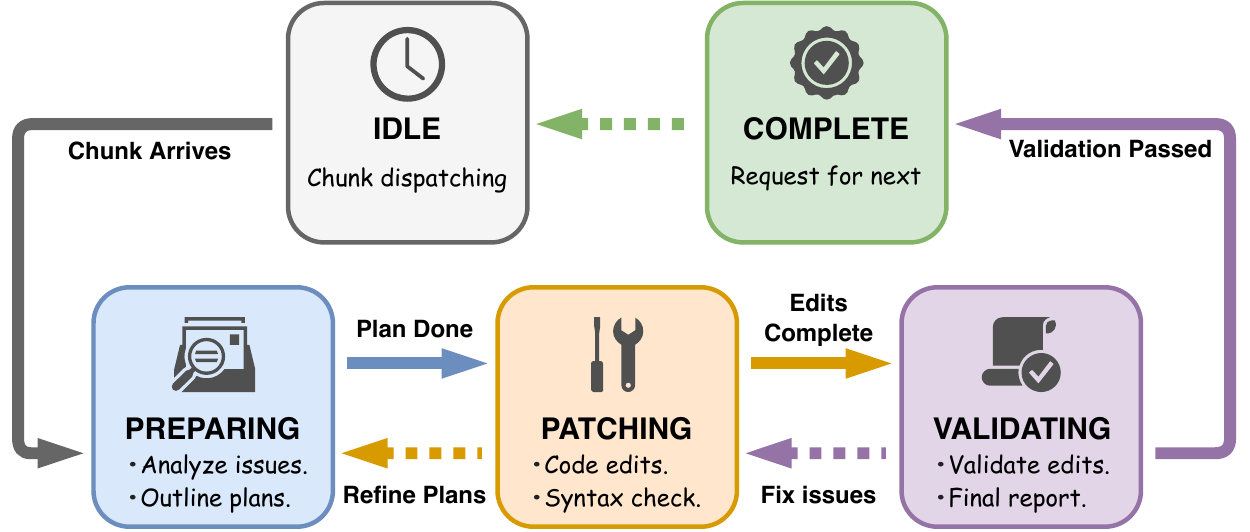}
  \caption{State machine workflow for the \textit{Patcher}.}
  \label{fig:state-machine}
\end{figure}

To make patching tractable, we first preprocess scanned results before passing them to the \textit{Patcher}. 
Duplicate findings are removed, and each target is expanded with related code context retrieved through language-server support, such as symbol references and associated locations. 
The resulting targets are then grouped into coherent chunks, typically at the source-file level, so that each patching task remains within the model context while preserving the relevant code context needed for coordinated edits.
In this way, preprocessing turns raw scan results into manageable and context-rich fixing units, reducing redundancy while providing enough surrounding information for stable patch generation.

\paragraph{State Machine.}

The transition from target preprocessing to patch generation is governed by a formalized multi-stage workflow encompassing planning, patching, and validating. 
This operational structure reflects the divergent requirements of each activity: planning necessitates comprehensive context acquisition, patching involves localized code transformation, and validation ensures the functional integrity of the resulting modifications. 
Allowing the agent to perform all of these activities in an unconstrained workflow can easily lead to drift, inconsistent edits, or syntax-breaking changes. 
To address this, the \textit{Patcher} is implemented as a structured state machine that constrains the agent’s behavior across stages.
By formalizing the boundaries between operational stages, this architecture restricts the agent's actions to contextually valid operations and provides explicit checkpoints for error recovery when the optimization process deviates from the intended plan.

\paragraph{Patching Workflow.}

As shown in Figure~\ref{fig:state-machine}, the workflow consists of three stages: \textit{Preparing}, \textit{Patching}, and \textit{Validating}. 
In \textit{Preparing}, the \textit{Patcher} analyzes the anti-pattern report and the current targets, gathers relevant context, and drafts a modification plan. 
In \textit{Patching}, it applies concrete code edits guided by that plan. 
In \textit{Validating}, it checks the modified code using language-server-based syntax analysis to ensure that no errors have been introduced and that the intended changes are complete.

Transitions between these stages are not strictly linear, allowing fallback on failure.
The agent normally progresses from preparation to editing and then to validation, while failures in later stages trigger a fallback to earlier ones. 
In particular, validation failures send the workflow back to patching, and insufficient context or infeasible edit plans trigger a fallback to preparation. 

Such backward transitions are accompanied by reverting to a previous checkpoint, preventing invalid intermediate edits from contaminating subsequent steps. 
By separating planning, editing, and validation in this way, \sys makes patch generation more stable, reliable and better suited to large-scale, context-dependent memory optimization.

After iterative preparation, patching, and validation, the \textit{Patcher} stably produces high-confidence optimization patches across the codebase. 
At this point, the mined anti-patterns are ultimately realized as concrete optimizations.

\section{Evaluation}
\label{sec:evaluation}

We explore the following research questions for \sys:

\begin{table*}[htbp]
\centering
\caption{Tools invoked by \sys}
\scriptsize
\label{tab:tools}
\begin{tabular}{@{}lp{0.30\textwidth}lp{0.30\textwidth}@{}}
\toprule
\textbf{Tool} & \textbf{Description} & \textbf{Tool} & \textbf{Description}\\
\midrule
\multicolumn{2}{@{}l}{\textbf{State machine control}} & \multicolumn{2}{l}{\textbf{Language Server (clangd)}}\\
\emph{update\_state} & Request a transition between states. & \emph{clangd\_hover/def/ref} & Return symbol info, definition, or refs.\\
\emph{show\_state} & Display current state and transitions. & \emph{clangd\_sync/diag} & Sync file content and show diagnostics.\\
\midrule
\multicolumn{2}{@{}l}{\textbf{Profiling data analysis}} & \multicolumn{2}{l}{\textbf{Workspace checkpoints}}\\
\emph{db\_list\_tables} & Show available tables and column info. & \emph{create\_checkpoint} & Create a checkpoint with description.\\
\emph{db\_describe\_table} & Describe table schema and show rows. & \emph{list\_checkpoints} & List all checkpoints and tracked files.\\
\emph{db\_query} & Run SQL against in-memory tables. & \emph{restore\_checkpoint} & Restore workspace to a specific point.\\
\midrule
\multicolumn{2}{@{}l}{\textbf{Task management}} & \multicolumn{2}{l}{\textbf{Checker synthesis}}\\
\emph{read\_todo} & Inspect todo list for active state. & \emph{write\_checker} & Write generated checker code.\\
\emph{write\_todo} & Add todo entries with priority/status. & \emph{build\_checker} & Run build commands for the checker.\\
\emph{update\_todo} & Update todo's priority/status/results. & \emph{test\_checker} & Run verification commands.\\
\midrule
\multicolumn{2}{@{}l}{\textbf{Code review \& editing}} & \multicolumn{2}{l}{\textbf{Reporting \& execution}}\\
\emph{list\_directory} & List files relative to project root. & \emph{review\_report} & Write report and invoke review.\\
\emph{read\_file} & Read a file with optional line range. & \emph{report} & Report the final result.\\
\emph{replace\_text} & Replace a literal string in a file. & \emph{run\_shell} & Execute shell command in project root.\\
\bottomrule
\end{tabular}
\end{table*}

\begin{itemize}[label=\large$\bullet$, leftmargin=2.5em]
    \item \textbf{RQ1}: How effectively can \sys identify actionable and previously unknown memory anti-patterns from profiling data?
    \item \textbf{RQ2}: To what extent can \sys synthesize static checkers with high diagnostic accuracy and practical utility?
    \item \textbf{RQ3}: To what extent can \sys generate effective optimization patches at repository scale?
    \item \textbf{RQ4}: How do the key components of \sys contribute to its overall effectiveness?
    \item \textbf{RQ5}: What are the resource costs of using \sys?
\end{itemize}

\subsection{Experimental Setup}
\label{sec:eval-setup}

We evaluate \sys under the following experimental setup.

\textit{Hardware and Software.}
Our experiments are conducted on a workstation with 24 cores, 32 GB RAM, running Ubuntu 22.04 LTS. 
We use LLVM 15.0.4 as the default compilation toolchain.

\textit{Subject System.}
We evaluate \sys on OpenHarmony 5.0, an open-source operating system comprising over 100 million lines of C/C++ code across dozens of system services~\cite{OpenHarmony}. 
The compilation configuration and test platform target the Rockchip RK3568 chip.

\textit{Profiling Setup.}
We use Memoro~\cite{byma2018memoro} for memory behavior profiling, which builds on the LLVM/Clang AddressSanitizer framework. 
The profiling data captures heap object sizes, lifecycles, and stack traces, and related runtime metadata.

\textit{LLM Configuration.}
We use OpenAI GPT-5.1 for pattern mining, and OpenAI GPT-5.1-Codex for checker synthesis and patch generation. 
The maximum iterations are set to 100 for pattern mining and checker synthesis, and 150 for patch generation per chunk. 
Table~\ref{tab:tools} summarizes the tools invoked by \sys across its three stages.

\subsection{RQ1: Pattern Mining Effectiveness}
\label{sec:eval-rq1}
\subsubsection{Overall Results}
As shown in Table~\ref{tab:pattern-categories}, \sys demonstrates strong capability in automated pattern discovery. 
From profiling data of just three system services, the framework successfully identified 13 validated anti-patterns, achieving a yield rate of 65\% from candidate reports. 
The low refinement overhead, averaging only 1.5 iterations per pattern, indicates that the Analyzer-Reviewer feedback loop converges efficiently without excessive back-and-forth. 
More importantly, when compared against Clang-Tidy's established performance checks, 69.3\% of our discovered anti-patterns have no existing coverage, confirming that profiling-guided mining can uncover blind spots missed by traditional rule-based approaches.

\begin{table}[t]
\centering
\caption{Categories of validated C/C++ memory anti-patterns.}
\label{tab:pattern-categories}
\scriptsize
\resizebox{0.97\columnwidth}{!}{%
    \begin{tabularx}{\columnwidth}{@{}lc>{\raggedright\arraybackslash}X>{\raggedright\arraybackslash}p{2.2cm}@{}}
    \toprule
    \textbf{Pattern Type} & \textbf{\#} & \textbf{Description} & \textbf{Impact} \\
    \midrule
    T1: Static object overuse & 4 & Excessive non-trivial static objects causing copies across translation units & Binary size bloat \\[0.5ex]
    T2: Inefficient strings & 2 & Frequent temporary string construction and concatenation & Allocation churn \\[0.5ex]
    T3: Redundant copying & 4 & Unnecessary implicit copies through pass-by-value and similar mechanisms & Copy overhead \\[0.5ex]
    T4: Const heap structures & 3 & Immutable complex structures persisting on heap throughout program lifetime & Persistent heap usage \\
    \midrule
    \textbf{Total} & \textbf{13} & & \\
    \bottomrule
    \end{tabularx}
}
\end{table}

\begin{table}[htbp]
\centering
\caption{Patterns overlap between \sys and Clang-Tidy.}
\label{tab:pattern-coverage}
\small
\begin{tabular}{ll}
\toprule
\textbf{\sys} & \textbf{Clang-Tidy} \\
\midrule
loop-local-string-concatenation & \texttt{inefficient-string-concatenation} \\
containers-pass-value-to-read-only-functions & \texttt{unnecessary-value-param} \\
pass-std-function-parameters-by-value & \texttt{unnecessary-value-param} \\
container-copy-using-auto-by-value & \texttt{unnecessary-copy-initialization} \\
\bottomrule
\end{tabular}
\end{table}

\subsubsection{Validated Patterns}
Table~\ref{tab:pattern-categories} presents the validated pattern categories with their descriptions and prevalence in the codebase. 
Overall, we categorize them into four types: T1 captures non-trivial static objects that might be replicated across translation units; T2 covers string temporaries and concatenations that drive allocation churn; T3 describes avoidable copy paths (e.g., pass-by-value) that incur extra copying work; and T4 focuses on long-lived immutable heap-resident data structures that unnecessarily occupy heap memory over the program’s lifetime.

Among the memory consumption related anti-patterns, \textit{Overuse of static object} (T1) is the most frequently occurring, accounting for 30.8\% of all detected instances. 
As exemplified in our motivating example (Figure~\ref{fig:motivating-example}), this pattern can cause significant binary size inflation and memory bloat due to redundant object copies and constructor invocations across translation units.

\subsubsection{Comparison with Clang-Tidy}

We compare \sys's mined anti-patterns against Clang-Tidy's built-in performance checks (\texttt{performance-*} rules), which represent the state-of-practice in rule-based detection.

\paragraph{Overlapping Patterns.} Table~\ref{tab:pattern-coverage} shows that 4 of our 13 anti-patterns (30.7\%) overlap with existing Clang-Tidy rules. 
These correspond to well-documented inefficiencies: loop-local string concatenation, unnecessary value parameters, and redundant copy initialization. 
The overlap validates that \sys can independently rediscover established anti-patterns from runtime evidence alone, without prior knowledge of existing rules.

\paragraph{Novel Patterns.} The remaining 9 anti-patterns (69.3\%) have no equivalent Clang-Tidy built-in check. 
For example, T1 (Static Object Overuse) causes significant binary bloat through redundant copies across translation units, yet no existing rule targets this issue. 
This gap arises because such anti-patterns manifest at link time or runtime rather than within a single translation unit, making them invisible to solely AST-based heuristics. 
The result highlights the value of profiling-guided discovery: runtime evidence reveals inefficiencies that static analysis alone cannot capture. 
We note that \sys and Clang-Tidy are complementary, the former surfaces emergent, project-specific anti-patterns while the latter provides fast detection of known issues with automatic fix-its. 

\begin{table}[htbp]
\centering
\caption{Detection results by checkers across OpenHarmony system services.}
\label{tab:scan-results-full}
\resizebox{0.93\textwidth}{!}{%
\begin{tabular}{lrrrrrrrrr}
\toprule
\textbf{Pattern} & \textbf{Render$^\dagger$} & \textbf{Camera$^\dagger$} & \textbf{Media} & \textbf{Audio$^\dagger$} & \textbf{AVSes.} & \textbf{Access.} & \textbf{Bgtask} & \textbf{Total} \\
\midrule
P1: Loop str concat        & \heatcell{33}    & \heatcell{6}     & \heatcell{3}     & \heatcell{1}     & \heatcell{27}   & \heatcell{0}      & \heatcell{0}    & \textbf{70}    \\
P2: Rebuilding str          & \heatcellL{2476}{2,476} & \heatcell{771}   & \heatcellL{1719}{1,719} & \heatcellL{1637}{1,637} & \heatcell{878}  & \heatcell{339}    & \heatcell{245}  & \textbf{8,065} \\
P3: Recopy container        & \heatcell{48}    & \heatcell{4}     & \heatcell{15}    & \heatcell{16}    & \heatcell{50}   & \heatcell{61}     & \heatcell{0}    & \textbf{194}   \\
P4: Non-constexpr str       & \heatcell{152}   & \heatcell{116}   & \heatcell{241}   & \heatcell{236}   & \heatcell{73}   & \heatcell{152}    & \heatcell{30}   & \textbf{1,000} \\
P5: Container copy auto     & \heatcell{52}    & \heatcell{6}     & \heatcell{5}     & \heatcell{0}     & \heatcell{5}    & \heatcell{2}      & \heatcell{0}    & \textbf{70}    \\
P6: Static vector tables    & \heatcell{50}    & \heatcell{40}    & \heatcell{38}    & \heatcell{35}    & \heatcell{6}    & \heatcell{3}      & \heatcell{0}    & \textbf{172}   \\
P7: Static map tables       & \heatcell{45}    & \heatcell{21}    & \heatcell{36}    & \heatcell{114}   & \heatcell{28}   & \heatcell{8}      & \heatcell{5}    & \textbf{257}   \\
P8: Static unordered map    & \heatcell{26}    & \heatcell{113}   & \heatcell{18}    & \heatcell{13}    & \heatcell{0}    & \heatcell{1}      & \heatcell{0}    & \textbf{171}   \\
P9: Static std::function    & \heatcell{7}     & \heatcell{0}     & \heatcell{0}     & \heatcell{0}     & \heatcell{0}    & \heatcell{0}      & \heatcell{0}    & \textbf{7}     \\
P10: Static set tables      & \heatcell{3}     & \heatcell{2}     & \heatcell{0}     & \heatcell{10}    & \heatcell{0}    & \heatcell{0}      & \heatcell{1}    & \textbf{16}    \\
P11: Pass std::func by val  & \heatcell{13}    & \heatcell{3}     & \heatcell{0}     & \heatcell{1}     & \heatcell{1}    & \heatcell{0}      & \heatcell{0}    & \textbf{18}    \\
P12: Containers pass val    & \heatcell{3}     & \heatcell{2}     & \heatcell{6}     & \heatcell{1}     & \heatcell{0}    & \heatcell{2}      & \heatcell{0}    & \textbf{14}    \\
P13: Per-instance str vec   & \heatcell{2}     & \heatcell{1}     & \heatcell{5}     & \heatcell{2}     & \heatcell{1}    & \heatcell{2}      & \heatcell{0}    & \textbf{13}    \\
\midrule
\textbf{Total}              & 2,910            & 1,085            & 2,086            & 2,066            & 1,069           & 570               & 281             & \textbf{10,067} \\
\bottomrule
\multicolumn{9}{l}{\small $^\dagger$ Services used for profiling in pattern mining.} \\
\end{tabular}
}%
\end{table}

\begin{table}[htbp]
\centering
\caption{Patch generation and optimization results for T1 \& T4.}
\label{tab:patch-by-pattern}
\resizebox{0.97\columnwidth}{!}{%
\newcommand{\fraction}[2]{\makebox[2em][c]{#1}/\makebox[2em][c]{#2}}
\begin{tabular}{lrrrrrrrr}
\toprule
\textbf{Service} & \textbf{LoC} & \textbf{Targets/Files} & \textbf{Patches} & \textbf{Valid} & \textbf{Acc} & \textbf{PSS$\downarrow$} & \textbf{Heap Size$\downarrow$} & \textbf{Binary Size$\downarrow$} \\
\midrule
Render Service      & 477~K & \fraction{431}{212}  & 322   & 286   & 88.8\%    & 6.3\%     & 39.4\%    & 23.6\%    \\
Camera Service      & 96~K  & \fraction{323}{51}   & 96    & 90    & 93.8\%    & 27.6\%    & 65.3\%    & 8.6\%     \\
Media Service       & 101~K & \fraction{352}{83}   & 109   & 99    & 90.8\%    & 25.1\%    & 48.0\%    & 15.1\%    \\
Audio Service       & 154~K & \fraction{412}{107}  & 136   & 123   & 90.4\%    & 11.9\%    & 40.7\%    & 15.1\%    \\
Avsession Service   & 49~K  & \fraction{97}{23}    & 44    & 43    & 97.7\%    & 23.8\%    & 53.0\%    & 6.4\%     \\
Accessibility       & 73~K  & \fraction{171}{27}   & 42    & 36    & 85.7\%    & 22.9\%    & 47.5\%    & 4.7\%     \\
Bgtaskmgr Service   & 18~K  & \fraction{39}{15}    & 20    & 20    & 100\%     & 1.4\%     & 1.4\%     & 0.5\%     \\
\midrule
\multicolumn{5}{l}{\textbf{Avg.}} & \textbf{92.5\%} & \textbf{17.0\%} & \textbf{42.2\%} & \textbf{10.6\%} \\
\bottomrule
\end{tabular}
}
\end{table}

\finding{1}{\sys successfully mines actionable memory anti-patterns from profiling data, with the majority representing novel issues not covered by existing static analysis tools. 
The automated validation effectively filters low-quality reports, demonstrating reliable and effective pattern discovery.}

\subsection{RQ2: Checker Effectiveness}
\label{sec:eval-rq2}

This research question evaluates how effectively synthesized checkers scale the detection of memory inefficiencies across the codebase.

From the 13 validated anti-patterns mined in RQ1, \sys successfully synthesized 13 corresponding checkers, all of which passed compilation and validation. 
We deploy these checkers to scan 7 representative OpenHarmony system services. 
Table~\ref{tab:scan-results-full} summarizes the scan outcomes.

The checkers detected a total of over 10,000 memory inefficiency instances across 7 system services with 971K lines of code. 
Notably, the anti-patterns are mined from profiling data collected on only 3 services (Render, Audio, and Camera), yet the checkers successfully identified issues in 4 additional services that were never profiled. 
This demonstrates that the synthesized checkers generalize beyond the original profiling scope, effectively scaling up the target locations for optimization. 

\finding{2}{The synthesized checkers accurately detect memory inefficiencies at scale and effectively generalize beyond the profiled services, enabling systematic propagation of pattern-based detection across the entire codebase without requiring additional profiling.}

\subsection{RQ3: Patch Generation and Optimization Effectiveness}
\label{sec:eval-rq3}

This research question evaluates whether \sys can generate effective optimization patches at scale and whether these patches deliver performance improvements in real-world deployments.

\subsubsection{Overall Results}

We select the scan results obtained in RQ2 from 7 checkers in pattern categories T1 and T4 across 7 core services and leverage \sys to generate optimization patches.

Table~\ref{tab:patch-by-pattern} presents the patch generation results. Across all seven services, \sys generates a total of 769 patches addressing 1,825 optimization targets distributed across 518 files. 
Here, we count patches by file—all modifications within a single file are treated as one patch. 
Notably, the framework achieves an average accuracy of 92.5\% in human validation, with 693 out of 769 patches deemed correct and ready for deployment. 
This high success rate is particularly significant for codebase-scale optimization efforts, as it substantially reduces the manual review burden that would otherwise be prohibitive when dealing with hundreds of patches across a large codebase.

An important observation is that the number of generated patches (769 files) exceeds the number of files containing targets (518 files), even though our preprocessing pipeline already merged multiple targets within each file. This discrepancy arises because many optimization targets require cross-file modifications. 
The ability of \sys to correctly identify and coordinate these inter-file dependencies demonstrates the effectiveness of our Patch Generation workflow design for repository-level code optimization. 

\subsubsection{Memory Impact}

To quantify the real-world impact of the optimizations, we measure three key metrics before and after deploying the validated patches: binary size, heap size, and proportional set size (PSS). 
Table~\ref{tab:patch-by-pattern} presents the results across all seven services, including both those used for initial profiling and those that were not.

The results strongly confirm our core insight: memory inefficiencies manifest as recurring anti-patterns rather than isolated anomalies. 
Notably, services that are \emph{not} included in the initial profiling phase still achieved substantial improvements. 
On average, the optimizations yield a 42.2\% reduction in heap size, 17.0\% reduction in PSS, and 10.6\% reduction in binary size, validating the scalability and effectiveness of our pattern-driven approach. 
This confirms our key insight that anti-patterns extracted from a small subset of profiled services can be systematically propagated to detect and repair similar inefficiencies across the entire codebase.

\subsubsection{Case Study}
During the pattern mining phase, we discover a critical memory bloat pattern in OpenHarmony's service layer (Figure~\ref{fig:motivating-example}): a constant HTTP status code mapping table (\texttt{std::map<int, std::string>}) defined in an anonymous namespace within a header file. 
Since this static non-trivial object resides in a header, the C++ compilation model requires each translation unit to instantiate its own initialization function and construct an independent copy at startup. 
Profiling reveals over one hundred duplicate instances across at least five services, each consuming heap memory for both the map structure and \texttt{std::string} values.

\begin{table}[htbp]
\centering
\caption{Patch Generation Results}
\small
\label{tab:patch_results}
\begin{tabular}{lccc}
\toprule
\textbf{Method} & \textbf{Total Patches} & \textbf{Successful} & \textbf{Success Rate} \\
\midrule
PatchAgent  & 165 & 17 & 10.3\% \\
MOA         & 322 (\textcolor{green!60!black}{$\uparrow$157})
            & 286 (\textcolor{green!60!black}{$\uparrow$269})
            & 88.8\% (\textcolor{green!60!black}{$\uparrow$78.5\%}) \\
\bottomrule
\end{tabular}
\end{table}

As shown in Figure~\ref{fig:motivating-example}, \sys proposes a solution that replaces the map-based lookup with a stateless function using a \texttt{switch} statement, and substitutes \texttt{std::string} objects with \texttt{const char*} pointers to string literals in the patch generation phase. 
This optimization eliminates all per-translation-unit initialization functions and heap allocations. 
Additionally, by applying synthesized checkers to systematically detect similar anti-patterns  and automatically optimizing them, \sys achieves substantial reductions in both memory consumption and binary size across multiple services. 

\subsubsection{Comparison with PatchAgent}

To demonstrate the effectiveness of our patch generation workflow design, we compare \sys's Patcher against PatchAgent~\cite{PatchAgent}, one of the most closely related LLM-based autonomous agent frameworks for program repair.
It reflects the current capability of LLMs in navigating complex codebases and generating functional patches through iterative tool-use.

While PatchAgent is originally designed for functional bug repair, we adapt it as a strong baseline for memory optimization, since its iterative code editing workflow is also applicable to repairing memory inefficiencies.
The original workflow of PatchAgent includes git diff patch generation, git apply, compilation, and vulnerability repair.
For a fair comparison in the OpenHarmony environment, where full compilation is prohibitively expensive, we retain PatchAgent's core git diff patch generation component and manually apply the resulting patches to the codebase.

Both frameworks are given the same buggy code path: PatchAgent uses its default retrieval database, whereas \sys uses pattern descriptions synthesized during the Pattern Mining stage. 
We evaluate both systems on a system service with over 200 memory inefficiencies detected by our checkers, and assess the generated patches for syntactic validity, semantic correctness through manual review, and expert acceptance.

As shown in Table~\ref{tab:patch_results}, \sys achieves an 88.8\% success rate compared to PatchAgent's 10.3\%, representing a 78.5 percentage point improvement. The advantage is particularly pronounced for complex optimizations: \sys successfully generates 43 multi-location patches that require coordinated changes across multiple code locations, which PatchAgent fails to attempt due to its single-shot generation approach.

We attribute \sys's superior performance to three key design choices in our workflow:
\begin{itemize}[label=\large$\bullet$, leftmargin=2.5em]
    \item \textit{Optimization guidance}: \sys provides the \textit{Patcher} with feasible fix strategies derived from validated anti-patterns.
    \item \textit{Context exploration}: \sys enables the agent to dynamically gather relevant code context through file operations and LSP queries, rather than relying on fixed retrieval that may miss critical dependencies.
    \item \textit{Iterative validation}: The LSP-based syntax checking allows the agent to detect errors early and refine patches through multiple attempts, whereas zero-shot generation lacks feedback mechanisms for self-correction.
\end{itemize}

PatchAgent excels in simplicity and speed—its zero-shot approach requires fewer LLM queries. For simple, single-location fixes, both tools perform comparably. 
However, for complex optimizations requiring contextual understanding, \sys's agent-based workflow (with state machine guidance, iterative context gathering, and LSP validation) is essential for effective large-scale memory optimization tasks and demonstrates clear benefits.

\finding{3}{\sys generates optimization patches at repository scale with high accuracy, achieving substantial memory improvements across multiple metrics. 
The framework successfully handles complex cross-file dependencies and coordinates multi-location modifications, demonstrating effectiveness for real-world deployment.}

\subsection{RQ4 \& RQ5: Ablation Study and Cost Analysis}
\label{sec:eval-rq4}

To understand the contribution of stages in \sys's pipeline, we conduct ablation studies by systematically removing key components. We also analyze the computational costs of \sys.

\subsubsection{Stage Contribution Analysis}

We evaluate two ablated configurations against the full \sys pipeline to isolate the impact of pattern guidance and precise localization. 

\begin{itemize}[label=\large$\bullet$, leftmargin=2.5em]
    \item \textbf{w/o Pattern Description}: The \textit{Patcher} receives only code locations flagged by checkers, without any description of the memory inefficiency or suggested fix strategy. This tests whether pattern mining provides essential guidance beyond mere localization.

    \item \textbf{w/o Checker Location}: The \textit{Patcher} receives only the pattern description (issue description and fix strategy) without checker-identified target locations. This tests whether the agent can locate optimization targets in a large codebase with only conceptual guidance.
\end{itemize}

\begin{table}[htbp]
\centering
\caption{Ablation study results on Camera Service.}
\label{tab:ablation-stages}
\small
\newcommand{\val}[2]{\makebox[2.5em][r]{#1}\makebox[3.5em][c]{#2}}
\begin{tabular}{lcccc}
\toprule
\textbf{Configuration} & \textbf{Patch} & \textbf{Acc} & \textbf{Binary$\downarrow$} & \textbf{PSS$\downarrow$} \\
\midrule
Full \sys                   & 96 & 93.8\%     & 8.6\%      & 27.3\%    \\
w/o Pattern Description     & 93 
                            & \val{91.4\%}{(\textcolor{red!60!black}{$\downarrow$2.4\%})}  
                            & \val{8.3\%}{(\textcolor{red!60!black}{$\downarrow$2.4\%})}   
                            & \val{27.2\%}{(\textcolor{red!60!black}{$\downarrow$0.2\%})} \\
w/o Checker Location        & 69 
                            & \val{47.8\%}{(\textcolor{red!60!black}{$\downarrow$46.0\%})} 
                            & \val{5.8\%}{(\textcolor{red!60!black}{$\downarrow$49.3\%})}  
                            & \val{25.3\%}{(\textcolor{red!60!black}{$\downarrow$2.8\%})} \\
\bottomrule
\end{tabular}
\end{table}

Table~\ref{tab:ablation-stages} shows the ablation study results.
The primary metric is the percentage of memory inefficiencies fixed relative to that achieved by the full \sys baseline.

We select Camera Service as the experimental target. 
Since optimizations across different OpenHarmony services can have cross-dependencies, we employ the following evaluation methodology to isolate the effects: we retain all optimizations for other services while applying different ablation configurations only to Camera Service. 
We define the optimization baseline as the full \sys approach, and calculate regression using:

\begin{equation}
    \text{Regression} = \frac{Opt_{\text{full}} - Opt_{\text{ablated}}}{Opt_{\text{full}}} \times 100\%
\end{equation}

where $Opt_{\text{full}}$ is the binary-size or PSS reduction achieved by the full \sys pipeline and $Opt_{\text{ablated}}$ is the corresponding reduction achieved by the ablated configuration, both computed from the underlying absolute measurements.
This metric quantifies how much optimization potential is lost when removing each component.

\paragraph{Impact of Pattern Mining.}
Removing anti-pattern descriptions results in modest degradation: accuracy drops by 2.4\%, while binary size reduction decreases slightly from 8.6\% to 8.3\% (2.4\% regression) and PSS reduction shows a negligible decline from 27.3\% to 27.2\% (0.2\% regression).

This relatively small impact suggests that modern LLMs have internalized common memory optimization strategies during pre-training. 
When presented with flagged code locations, the model can often infer the underlying issue and apply appropriate fixes based on its learned knowledge. However, the presence of explicit pattern descriptions still provides measurable value by reducing ambiguity and improving patch quality.

\paragraph{Impact of Checker Synthesis.}
In stark contrast, removing checker-provided localization causes severe degradation: accuracy plummets to 47.8\% (46.0\% drop), while binary size reduction falls to 5.8\% (49.3\% regression) and PSS reduction decreases to 25.3\% (2.8\% regression). 
This dramatic decline reveals a critical limitation: even when armed with precise pattern descriptions, the LLM-based agent struggles to autonomously locate optimization targets within a large codebase. 
The codebase scale overwhelms the agent's ability to systematically search and identify all relevant code instances matching the pattern.

This finding validates Checker Synthesis as a pivotal bridge between Pattern Mining and Patch Generation. 
Automated checkers facilitate exhaustive and precise localization across the codebase, thereby allowing the agent to dedicate more of its reasoning resources to generating correct fixes.

\subsubsection{Cost Analysis}
Table~\ref{tab:cost-breakdown} presents the cost breakdown by stage in terms of token consumption, monetary cost, and wall-clock time.

\begin{table}[t]
\centering
\caption{Cost breakdown by stage (average per instance).}
\small
\label{tab:cost-breakdown}
\begin{tabular}{lrrr}
\toprule
\textbf{Stage} & \textbf{Tokens} & \textbf{Cost (\$)} & \textbf{Time} \\
\midrule
Pattern Mining (per pattern)    & 1,115,242 & \$1.84    & 12.1~min  \\
Checker Synthesis (per checker) & 620,150   & \$1.49    & 14.5~min  \\
Patch Generation (per patch)    & 176,689   & \$0.50    & 7.8~min   \\
\bottomrule
\end{tabular}
\end{table}

As shown in Table~\ref{tab:cost-breakdown}, the \textit{Analyzer} costs \$1.84 per pattern and takes 12.1 minutes on average, the \textit{Checker Generator} costs \$1.49 per checker with 14.5 minutes of processing time on average, and the \textit{Patcher} costs \$0.50 per patch with 7.8 minutes on average. 
For our entire evaluation pipeline across all services, the total cost of \sys remains under \$500 overall.

These costs demonstrate that \sys operates within acceptable and manageable resource constraints for codebase-scale optimization. 
The modest per-patch cost and reasonable processing time make it practical to apply \sys across large codebases.

\finding{4}{All the key components are essential to \sys's effectiveness. Ablation studies confirm that pattern guidance improves patch quality while checker-based localization is critical for scaling optimization across large codebases. The overall pipeline operates within practical cost constraints suitable for industrial adoption.}

\section{Discussion}
\label{sec:discussion}

\subsection{Limitations}

While \sys demonstrates the viability of LLM-driven automation across pattern mining, checker synthesis, and repair generation, we acknowledge several limitations and open challenges that warrant further investigation.

\paragraph{Programming Language Support.}
\sys currently supports only C/C++ codebases, limiting its direct applicability to projects in other languages such as Java, Python, or JavaScript. 
However, the modular architecture is designed for extensibility. 
Supporting new languages requires adapting language-specific profiling methods, static analyzers and Language Server, without fundamental changes to the core pipeline.

\paragraph{Performance Metrics.}
Our evaluation mainly focuses on memory-related performance inefficiencies identified through Memoro profiling. 
While this addresses critical concerns in embedded systems like OpenHarmony, \sys does not yet directly handle other performance dimensions such as CPU efficiency, cache behavior, or I/O latency. 
Nonetheless, the SQL-based pattern mining can readily accommodate alternative profiling tools and metrics, provided the data maintains sufficient granularity and structure.

\subsection{Future Directions}

Looking forward, we identify several promising directions for extending \sys's capabilities and broadening its impact.

\paragraph{Functional Optimizations.}
\sys currently targets non-functional inefficiencies that preserve program semantics, such as redundant allocations and suboptimal data structures, and has demonstrated substantial optimization gains in this domain. 
However, we believe significant optimization potential also exists in functional aspects of code design, such as architectural refactoring, algorithm selection, and computational complexity reduction. 
Addressing these opportunities requires deeper semantic reasoning and more sophisticated validation mechanisms to ensure correctness during complex transformations. 
Future work will explore extending \sys to safely handle such functional optimizations.

\paragraph{Project Maintainability.}
\sys optimizes for performance metrics without explicitly considering the impact on code maintainability. 
Certain optimizations may trade readability or flexibility for performance gains. 
Future work should incorporate maintainability metrics (cyclomatic complexity, coupling, cohesion) to enable balanced trade-offs, potentially offering developers multiple patches with different performance-maintainability profiles aligned with project-specific priorities.
\section{Related Works}

\paragraph{\textbf{Performance Bug Analysis and Detection}}
Performance bugs are prevalent in large-scale software systems and differ significantly from functional bugs. 
As highlighted in a comprehensive study~\cite{jin2012understanding}, performance inefficiencies often do not cause crashes but lead to significant resource inefficiencies, and they frequently stem from recurring suboptimal coding patterns~\cite{han2018perflearner}. 
Traditionally, developers rely on dynamic profiling tools such as Gprof~\cite{Gprof}, gperftools~\cite{gperftools} and Valgrind~\cite{Valgrind} to pinpoint performance bottlenecks~\cite{10.1145/2345156.2254074}. 
On the other hand, static analysis tools like Clang Static Analyzer~\cite{ClangStaticAnalyzer} and Infer~\cite{Infer} offer better coverage but require experts to manually formalize code patterns into checkers. 
\sys bridges this gap by serving dynamic traces as foundation to automatically synthesize static checkers via LLMs, combining the precision of dynamic evidence with the coverage of static analysis.

\paragraph{\textbf{LLM-based Bug Detection}}
Recent research has explored using Large Language Models (LLMs) to automate the creation of static analysis rules. 
KNighter~\cite{KNighter} represents a significant step in this direction by synthesizing Clang Static Analyzer checkers from documentation and natural language specifications. 
Several recent works apply LLMs to vulnerability detection and rule synthesis. 
IRIS~\cite{li2025iris} couples LLM inference with repository-wide static reasoning by automatically inferring taint specifications, substantially improving vulnerability recall over CodeQL~\cite{CodeQL} baselines on Java benchmarks. 
QLCoder~\cite{Wang2025QLCoder} focuses on query synthesis, generating CodeQL queries directly from CVE metadata through an agentic loop with execution feedback and tooling support. 
QLPro~\cite{Hu2025QLPro} combines LLM reasoning with static analysis outputs and multi-role voting mechanisms for vulnerability discovery across open source projects. 
In supply chain security, RuleLLM~\cite{zhang2025automatically} uses LLMs to generate YARA~\cite{YARA} and Semgrep~\cite{Semgrep} rules for detecting malicious packages from code and metadata.
While these approaches focus on security vulnerabilities or general bug detection from documentation, \sys targets the system-level memory optimization.

\paragraph{\textbf{Autonomous LLM Agents for Program Repair}}
Automated Program Repair (APR) has transitioned from traditional heuristic-based approaches to LLM-driven conversational and agentic frameworks~\cite{Zhang2025-kb, Yang2025-ti, ZUBAIR2025103951}. 
ChatRepair~\cite{ChatRepair} demonstrates the effectiveness of iterative conversations with LLMs to fix functional bugs cost-effectively. 
To handle more complex reasoning, ThinkRepair~\cite{ThinkRepair} introduces a self-directed "thought" process, allowing the model to reason about the bug before generating patches. 
More recently, the field has moved towards autonomous agents that mimic human debugging workflows. 
RepairAgent~\cite{RepairAgent} and PatchAgent~\cite{PatchAgent} utilize specialized tools and feedback loops to autonomously navigate the codebase, gather context, and validate patches. 

\paragraph{\textbf{LLM-based Performance Optimization}}
A growing body of work focuses specifically on using LLMs to improve code efficiency~\cite{Zhang2023-em, sheng2025llms}. 
RAPGen~\cite{RAPGen} investigates the zero-shot capabilities of LLMs in fixing code inefficiencies, showing that LLMs can identify suboptimal patterns without extensive training. 
In the context of large-scale infrastructure, ECO~\cite{ECO} provides an LLM-driven optimizer tailored for warehouse-scale computers to reduce resource consumption. 
Other works focus on specific domains or methodologies; for instance, XRFix~\cite{XRFix} explores performance inefficiencies repair specifically for Extended Reality (XR) applications, while SemOpt~\cite{SemOpt} combines LLMs with rule-based analysis to drive optimizations. 
These approaches largely rely on \emph{repository-history mining} to distill optimization patterns from historical commits. 
However, this paradigm is bounded by previously identified fixes and often misses latent inefficiencies without prior PR records~\cite{bird2009fair, guo2010characterizing}. 
Conversely, \sys utilizes \emph{runtime profiling evidence} to expose concrete symptoms, enabling us to uncover and generalize memory anti-patterns that remain undiscovered in the codebase.

\section{Conclusion}

In this paper, we introduced \sys, an LLM-driven framework that bridges the gap between localized profiling symptoms and codebase-wide memory optimization. 
By combining pattern mining, checker synthesis, and patch generation, \sys automates the full workflow from runtime evidence to deployed fixes. 
An evaluation on OpenHarmony demonstrates that \sys identifies 13 anti-patterns, of which 69.3\% are not covered by existing Clang-Tidy rules, detects over 10,000 instances across the codebase, and generates 769 patches with a 92.5\% acceptance rate, yielding 42.2\% heap reduction and 10.6\% binary size reduction on average. 
With reasonable costs averaging \$0.50 per file, \sys holds strong potential as a practical tool for developers and researchers seeking to optimize memory across large-scale software systems.

\section{Data Availability Statement}

The artifact of this paper is publicly available at \url{https://doi.org/10.5281/zenodo.19248417}.




\end{document}